# $\beta$-$(Al_{0.19}Ga_{0.81})_2O_3$/$\beta$-$Ga_2O_3$ Modulation-Doped Field-Effect Transistors with > 6 kV Breakdown

Julian Gervassi-Saga, Joshua T. Buontempo, Nabasindhu Das, Advait Gilankar, Hari P. Nair, and Nidhin Kurian Kalarickal

***Abstract*— $\beta$-$(Al_{0.19}Ga_{0.81})_2O_3$/$\beta$-$Ga_2O_3$ modulation-doped field-effect transistors (MODFETs) incorporating a high-quality $Al_2O_3$ gate dielectric and $SiN_x$ passivation are demonstrated for high-voltage operation. The devices exhibited a maximum drain current of 54 mA/mm, an on-resistance of 149 Ω·mm, a minimum subthreshold slope of 94 mV/dec, and an $I_{ON}/I_{OFF}$ ratio exceeding $10^8$. Three-terminal breakdown measurements performed in Flourinert demonstrated a breakdown voltage that increased with gate-drain spacing, reaching 6.6 kV for a device with $L_{GD}$ = 28 µm. The corresponding average lateral electric field approached 2.4 MV/cm, while a power figure of merit of 222 $MW/cm^2$ was achieved. To the best of the authors' knowledge, the demonstrated breakdown voltage represents the highest reported value for a $\beta$-$(Al_xGa_{1-x})_2O_3$/$\beta$-$Ga_2O_3$ MODFET and highlights the potential of modulation-doped $\beta$-$Ga_2O_3$ heterostructures for next-generation multi-kV power devices.**

## I. INTRODUCTION

MONOCLINIC $\beta$-$Ga_2O_3$ exhibits an ultrawide bandgap (~4.8 eV) and correspondingly high, estimated, critical electric field (~8 MV/cm) [1], [2], making it an attractive platform for high-voltage and high-power switching applications [3], [4]. Lateral field-effect transistors provide a promising solution for power integration, primarily due to the compatibility with planar processing [5], [6]. Prior work on lateral $\beta$-$Ga_2O_3$ transistors has primarily focused on uniformly doped channels and has demonstrated breakdown voltages exceeding 10 kV and power figures of merit reaching up to 1 $GW/cm^2$ [7], [8]. However, such channels are not ideal for power transistors since lightly doped channels are required to achieve high mobility, which in-turn requires thick channels (> 300 nm) for reasonable channel conductivity, hence large gate lengths (> 3 microns) to prevent short channel effects such as punch through [4], [9], [10], [11], [12], [13]. This tradeoff can be mitigated using $\beta$-$(Al_xGa_{1-x})_2O_3$/$\beta$-$Ga_2O_3$ modulation-doped heterostructures. Due to the lack of ionized impurity scattering, a $\beta$-$(Al_xGa_{1-x})_2O_3$/$\beta$-$Ga_2O_3$ two-dimensional electron gas (2DEG) can simultaneously achieve high channel mobility and channel conductivity. Furthermore, since 2DEG channels can be placed close to the gate, smaller gate lengths can be employed with minimal short channel effects.

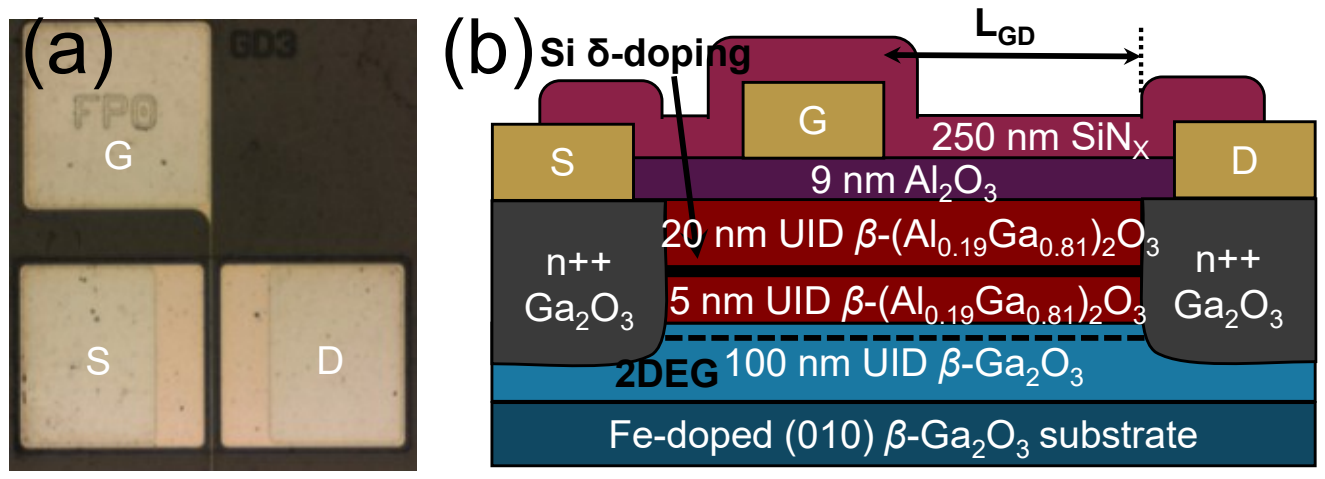

Fig. 1. (a) Optical image and (b) two-dimensional schematic of the fabricated $\beta$-$(Al_{0.19}Ga_{0.81})_2O_3$/$\beta$-$Ga_2O_3$ MODFET with dimensions $L_{GS}/L_G/L_{GD}$=1.5/1/3 µm.

In addition to channel engineering, field management is critical to achieving high breakdown voltages. Under high drain bias, electric field crowding near the gate edge can induce premature breakdown well below the critical field of $\beta$-$Ga_2O_3$. To mitigate field crowding, field plates, hetero-superjunctions [14], high-k or extreme-k dielectrics have been investigated to redistribute the electric field. Such approaches have been studied in $\beta$-$Ga_2O_3$ MESFET and MOSFET architectures [15], [16], [17], [18], [19], [20], [21], [22], [23], [24], [25]. To date, reported $\beta$-$(Al_xGa_{1-x})_2O_3$/$\beta$-$Ga_2O_3$ modulation-doped transistors have primarily focused on radio-frequency performance [26], [27], [28], [29], [30], [31], [32], [33], and the potential of this heterostructure for high-voltage operation remains largely unexplored [34], [35], [36].

In this work, we report lateral $\beta$-$(Al_{0.19}Ga_{0.81})_2O_3$/$\beta$-$Ga_2O_3$ modulation-doped field-effect transistors incorporating an $Al_2O_3$ gate dielectric and $SiN_x$ passivation achieving a breakdown voltage exceeding 6 kV. To the best of the authors' knowledge, this represents the highest reported breakdown voltage for a $\beta$-$(Al_xGa_{1-x})_2O_3$/$\beta$-$Ga_2O_3$ modulation-doped transistor and ranks among the highest for lateral $\beta$-$Ga_2O_3$ transistors.

## II. DEVICE GROWTH AND FABRICATION

The $\beta$-$(Al_xGa_{1-x})_2O_3$/$\beta$-$Ga_2O_3$ heterostructures were grown by metal-organic chemical vapor deposition MOCVD in a cold-wall Agnitron Agilis 100 reactor on Fe-doped (010) $\beta$-$Ga_2O_3$ substrates. After an *ex situ* HF clean, a ~50 nm UID $\beta$-$Ga_2O_3$ buffer was grown at 600 °C and 15 Torr using a TEGa molar flow of ~27 µmol/min and an $O_2$ flow of 500 sccm. The substrate temperature was then increased to 650 °C and the

Manuscript submitted for review August 19, 2026.
This material is based upon work supported by Army Research Office UWBG RF center under award No. W911NF2520005 (managed by Dr. Tom Oder). The use of the core facilities at ASU is supported, in part, by NSF under Award ECCS-2025490.
Julian Gervassi-Saga, Advait Gilankar, Nabasindhu Das, and Nidhin Kurian Kalarickal are with the School of Electrical, Computer, and Energy Engineering, Arizona State University, Tempe, AZ 85287 USA (e-mail: jgervass@asu.edu)
Joshua T. Buontempo and Hari P. Nair are with the Department of Material Science and Engineering, Cornell University, Ithaca, NY 14583 USA.

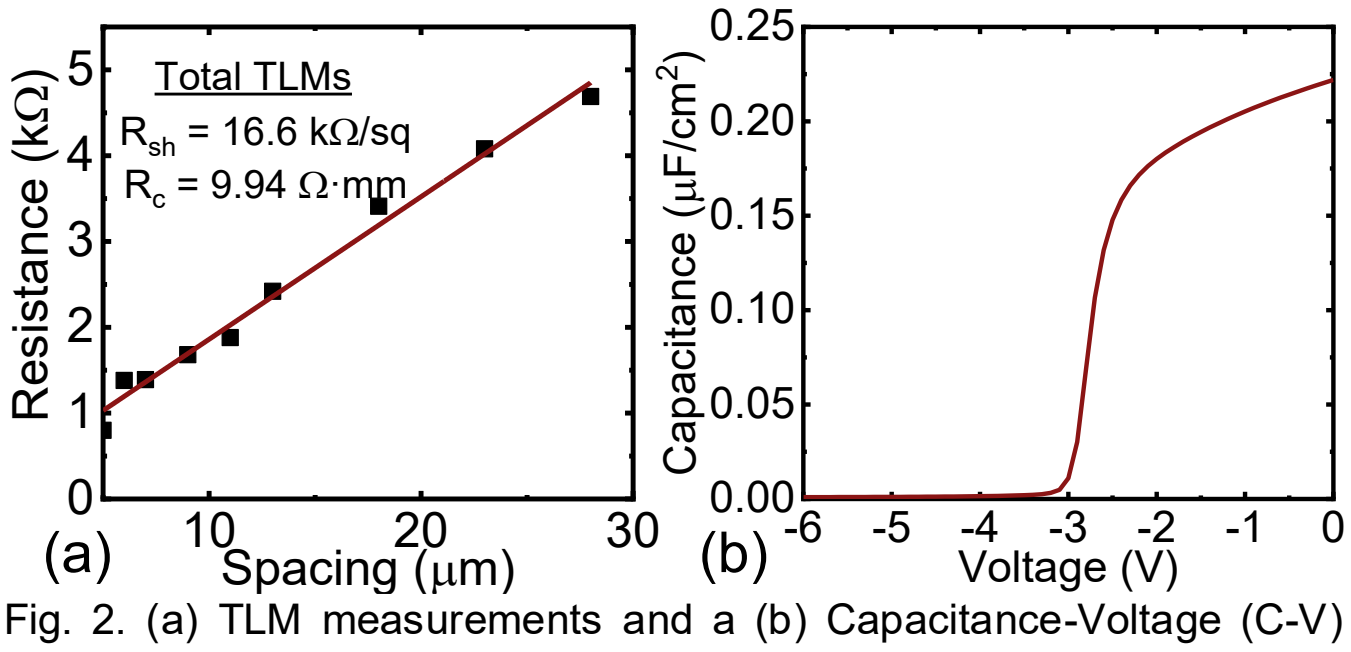

Fig. 2. (a) TLM measurements and a (b) Capacitance-Voltage (C-V) measurement probing the channel.

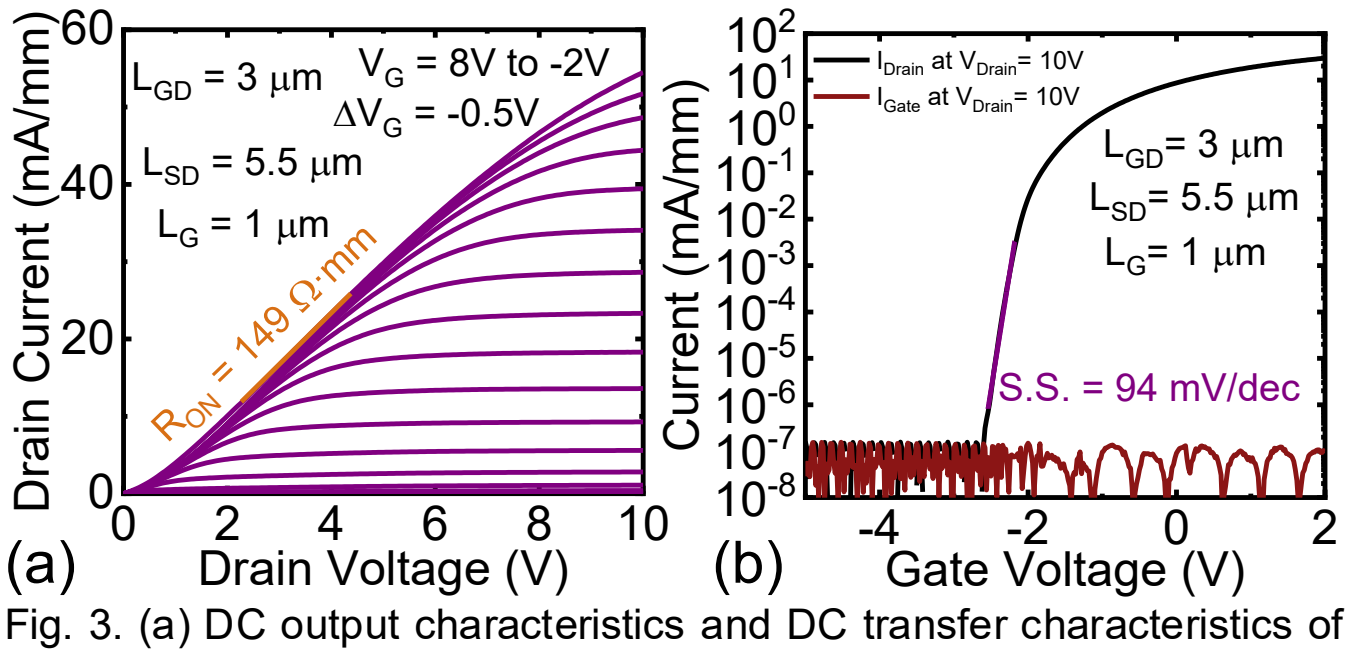

Fig. 3. (a) DC output characteristics and DC transfer characteristics of the MODFET with dimensions $L_{GS}/L_G/L_{GD}$=1.5/1/3 μm.

reactor pressure to 40 Torr, where an additional ~50 nm UID $\beta$-$Ga_2O_3$ buffer was grown using ~38 μmol/min of TEGa and 200 sccm of $O_2$. A ~5 nm $\beta$-($Al_{0.19}Ga_{0.81}$)$_2O_3$ spacer was subsequently grown using ~4.2 μmol/min of TEAl and ~34 μmol/min of TEGa. Following a 20 s precursor purge, a 1 nmol Si δ-doping dose was supplied using 25 ppm $SiH_4$ balanced in Ar. After a second 20 s purge, a ~20 nm $\beta$-($Al_{0.19}Ga_{0.81}$)$_2O_3$ barrier was grown.

The source and drain regions were defined using 500 nm thick PECVD $SiO_2$ hard mask. Following patterning, an 80-nm recess etch was performed to remove the $\beta$-$Ga_2O_3$ cap and $\beta$-($Al_{0.19}Ga_{0.81}$)$_2O_3$ barrier layers to expose the $\beta$-$Ga_2O_3$ channel. The recess etch was carried out using a low-damage $BCl_3$-based ICP-RIE process ($BCl_3$ flow = 20 sccm, ICP power = 20 W, RF power = 15 W, pressure = 5 mTorr). A heavily Si-doped, 120 nm thick $n^{++}$ $\beta$-$Ga_2O_3$ layer was regrown by MOCVD and lifted off in BOE [37]. Hall measurements revealed an electron concentration and mobility of $1.4 \times 10^{20}$ $cm^{-3}$ and 72 $cm^2$/Vs, respectively for the regrown contact layer. A Ti/Ni (30 nm/100 nm) stack was subsequently deposited onto the regrown $\beta$-$Ga_2O_3$ surface using electron-beam evaporation, followed by a rapid-thermal anneal (470˚C, 1 min, $N_2$ ambient). Transmission line measurements (TLM) on the regrown $n^{++}$ $\beta$-$Ga_2O_3$ yielded a sheet resistance of 57 Ω/sq and a metal/regrowth contact resistance of 0.12 Ω·mm. Mesa isolation of the active regions was performed using a $BCl_3$/Ar-based ICP-RIE process. To suppress gate leakage, 9 nm thick $Al_2O_3$ layer was deposited by plasma-assisted atomic layer deposition. A 120 nm thick Ni gate was deposited via electron-beam evaporation. Finally, the devices were passivated using 250 nm $SiN_x$ deposited via PECVD at 350 °C. The device structure is shown in Fig. 1 (b), consisting of a fixed gate to source spacing ($L_{GS}$) of 1.5 μm, a gate length ($L_G$) of 2 μm, and device width of 100 μm. The gate to drain distance ($L_{GD}$) varied from 6 μm to 28 μm. Smaller devices with dimensions $L_{GS}/L_G/L_{GD}$=1.5/1/1.5 μm and $L_{GS}/L_G/L_{GD}$=1.5/1/3 μm were also fabricated.

## III. Results and Discussions

Hall measurements yielded a sheet carrier density, hall mobility, and sheet resistance of $3.2 \times 10^{12}$ $cm^{-2}$, 132 $cm^2$/Vs, and 14.8 kΩ/sq, respectively for the $\beta$-($Al_{0.19}Ga_{0.81}$)$_2O_3$/$\beta$-$Ga_2O_3$ channel. The 2DEG mobility is below that of prior reports [38], likely due to the presence of a parallel conduction in the $\beta$-($Al_{0.19}Ga_{0.81}$)$_2O_3$ delta-doped layer. Capacitance-voltage (C-V) measurements at 100 kHz revealed a total 2DEG sheet charge density of $3.3 \times 10^{12}$ $cm^{-2}$. The C-V profile shows the absence of parasitic channels at the substrate/epi interface.

Fig. 3 (a), (b) show the DC output and transfer characteristics of the device with dimensions $L_{GS}/L_G/L_{GD}$=1.5/2/3 μm. The transfer characteristics exhibited a threshold voltage of −0.7 V, using a 1 mA/mm drain current. The devices demonstrated a subthreshold slope of 94 mV/dec, $I_{ON}/I_{OFF}$ ratio exceeding $10^8$ and an off-state leakage current below $1.5 \times 10^{-10}$ A/mm. The transconductance characteristics exhibited a peak $g_m$ of approximately 11 mS/mm at $V_{DS}$ = 10 V. The output characteristics yielded a max drain current ($I_D^{max}$) of 54 mA/mm and an on-resistance ($R_{ON}$) of 149 Ω·mm. From TLM measurements in Fig. 2 (a), the total contact resistance ($R_c^{total}$) was measured to be 9.94 Ω·mm (metal to channel). TLM measurements indicate that the dominant contribution to $R_c^{total}$ originates from the regrowth/channel interfacial resistance ($R_{int}$). A channel sheet resistance of 16.6 kΩ/sq was also extracted from the TLM structures. In comparison, a channel sheet resistance of 17.4 kΩ/sq and an intercept of 71.6 Ω·mm were extrapolated by linear fitting of $R_{ON}$ measured from devices as a function of $L_{SD}$. The intercept corresponds to the parasitic resistances that are independent of $L_{GD}$ such as the contact resistance, source-gate resistance ($R_{SG}$), and channel resistance under the gate ($R_G$). The parasitic resistance and sheet resistance agrees closely with the values from TLM measurements ($R_{sh}$=16.6 kΩ/sq, $R_{parasitic}$= $2R_c + R_{SG} + R_G$ = 71.57 Ω·mm).

Fig. 4 (a) shows the breakdown characteristics of the MODFETs, measured in Flourinert for various $L_{GD}$. The breakdown voltage increased monotonically with $L_{GD}$, ranging from 545 V for $L_{GD}$ = 3 μm to 6.6 kV for $L_{GD}$ = 28 μm. The highest measured breakdown voltage of 6.6 kV is among the highest reported for lateral $Ga_2O_3$ FETs and represents the highest breakdown voltage reported for $\beta$-($Al_xGa_{1-x}$)$_2O_3$/$\beta$-$Ga_2O_3$ MODFETs. The corresponding average breakdown field, calculated as ($E_{br} = V_{br}/L_{GD}$), showed a slight increase with increasing $L_{GD}$ and saturated at ~2.4 MV/cm for devices with ($L_{GD}$ > 13 μm). To examine the field distribution at breakdown, 2D electrostatic simulations were performed for the device with $L_{GD}$ = 13 μm, biased at its respective experimental breakdown voltage of $V_{DS}$ = 3 kV. The gate geometry incorporated a 10 nm gate-edge radius to represent finite corner curvature. Fig. 5 (a) shows that the field is strongly localized near the gate edge as expected in the absence of field-management structures, with a peak electric field of ~12 MV/cm in the $\beta$-($Al_{0.19}Ga_{0.81}$)$_2O_3$ cap layer.

The specific on-resistance ($R_{ON}^{sp}$) was obtained by normalizing the measured $R_{ON}$ with respect to $L_{SD}$. The power

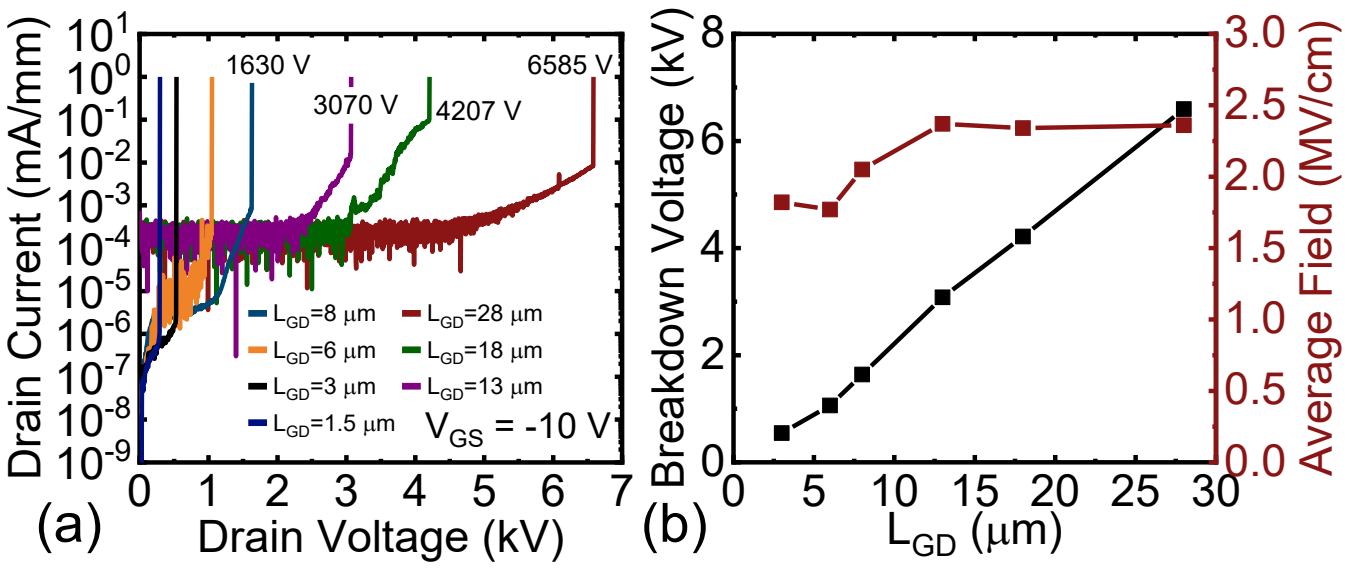

Fig. 4. (a) Reverse leakage current as a function of drain voltage for transistors with varying $L_{GD}$. (b) Breakdown voltage and corresponding average electric field as a function of $L_{GD}$.

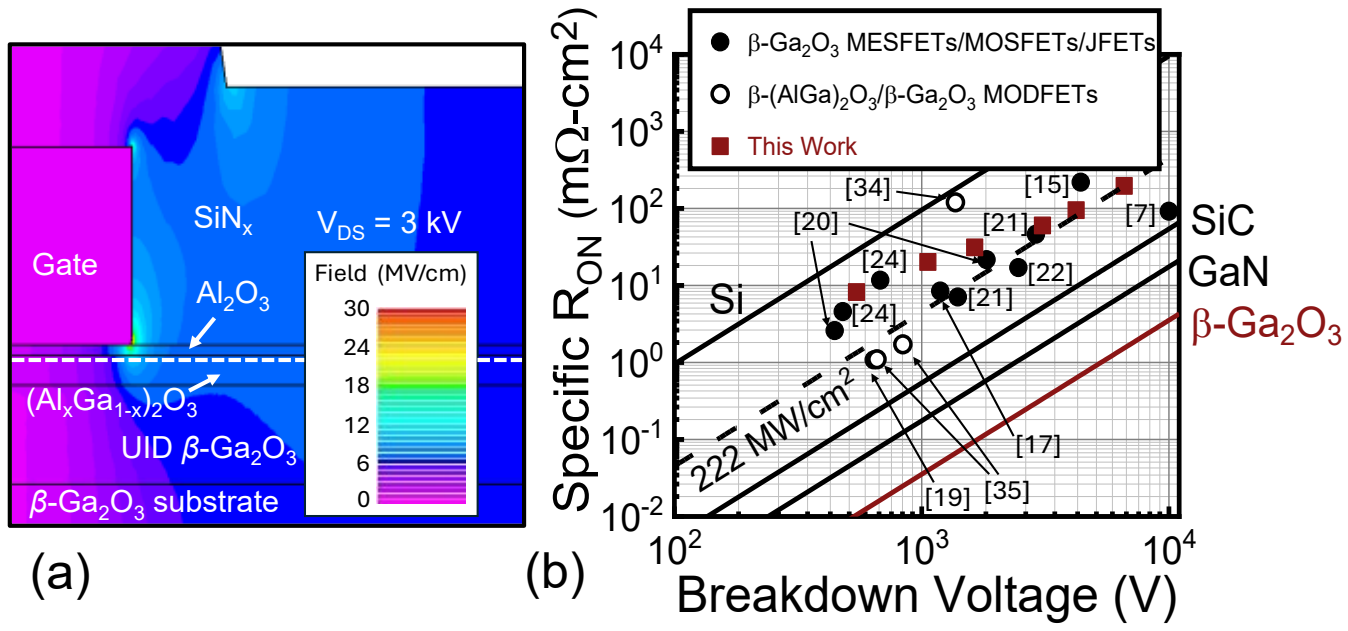

Fig. 5. (a) Simulated electric field contour profile for the MODFET with dimensions $L_{GS}/L_G/L_{GD}$=1.5/2/13 µm at breakdown voltage. (b) $R_{ON}$ - $V_{br}$ benchmark plot for lateral $\beta$-$Ga_2O_3$ transistors.

figure of merit (PFOM) was subsequently calculated using PFOM = $V_{br}^2$ / $R_{ON}^{sp}$. The device with ($L_{GD}$ = 3 µm) exhibited an $R_{ON}^{sp}$ of 8.2 mΩ·cm², a breakdown voltage of 545 V, and a PFOM of 35 MW/cm². In contrast, the device with ($L_{GD}$ = 28 µm) achieved an $R_{ON}^{sp}$ of 196 mΩ·cm², a breakdown voltage of 6.6 kV, and a PFOM of 222 MW/cm² which is the highest reported for $\beta$-($Al_xGa_{1-x}$)$_2O_3$/$\beta$-$Ga_2O_3$ MODFETs. Fig. 5 (b) benchmarks the devices with previously reported $\beta$-$Ga_2O_3$ power transistors in terms of PFOM.

## IV. Conclusion

In conclusion, we demonstrate $\beta$-($Al_{0.19}Ga_{0.81}$)$_2O_3$/$\beta$-$Ga_2O_3$ MODFETs with a maximum three-terminal breakdown voltage of 6.6 kV and a power figure of merit of 222 MW/cm², establishing the highest reported breakdown voltage for a $\beta$-($Al_xGa_{1-x}$)$_2O_3$/$\beta$-$Ga_2O_3$ MODFET. The ability to sustain average lateral fields of 2.4 MV/cm without field plates or edge termination suggests substantial opportunity for further performance improvement through advanced field management techniques. These results demonstrate that modulation-doped $\beta$-($Al_xGa_{1-x}$)$_2O_3$/$\beta$-$Ga_2O_3$ heterostructures are a promising platform for high-voltage power electronics.

## Acknowledgment

The use of the core facilities at ASU is supported, in part, by NSF under Award ECCS-2025490. This work utilized the Cornell NanoScale Facility, a member of the National Nanotechnology Coordinated Infrastructure (NNCI), which is supported by the NSF (Grant No. NNCI-2025233).